\documentclass[aps,prl,preprint,groupaddress]{revtex4}

\usepackage{bbm}
\usepackage{graphicx}
\usepackage{subfigure}
\usepackage{amsmath}
\usepackage{booktabs}

\begin{document}

\title{Possible Meissner effect near room temperature in copper-substituted lead apatite}

\author{Hongyang Wang$^{1}$\footnote{\url{wanghy@ipe.ac.cn}}, Yao Yao$^{2}$\footnote{\url{yaoyao2016@scut.edu.cn}}, Ke Shi$^{3}$, Yijing Zhao$^{3}$, Hao Wu$^{4}$, Zhixing Wu$^{5}$, Zhihui Geng$^{6}$, Shufeng Ye$^{1}$, and Ning Chen$^{7}$}

\address{$^1$ Center of Materials Science and Optoelectronics Engineering, Institute of Process Engineering, Chinese Academy of Sciences, Beijing 100049, China\\
$^2$ State Key Laboratory of Luminescent Materials and Devices and Department of Physics, South China University of Technology, Guangzhou 510640, China\\
$^3$ Beijing 2060 Technology Co., Ltd, Beijing 100084, China\\
$^4$ School of Materials Science and Engineering, Huazhong University of Science and Technology, Wuhan 430074, China\\
$^5$ Fujian Provincial Key Laboratory of Analysis and Detection Technology for Food safety, College of Chemistry, Fuzhou University, Fuzhou 350108, China\\
$^6$ School of Engineering, Course of Applied Science, Tokai University, Hiratsuka 2591292, Japan\\
$^7$ School of Materials Science and Engineering, University of Science and Technology Beijing, Beijing 100083, China}

\date{\today}

\begin{abstract}
With copper-substituted lead apatite below room temperature, we observe diamagnetic dc magnetization under magnetic field of 25~Oe with remarkable bifurcation between zero-field-cooling and field-cooling measurements, and under 200~Oe it changes to be paramagnetism. A glassy memory effect is found during cooling. Typical hysteresis loops for superconductors are detected below 250~K, along with an asymmetry between forward and backward sweep of magnetic field. Our experiment suggests at room temperature the Meissner effect is possibly present in this material.
\end{abstract}

\maketitle

Perfect diamagnetism, namely the Meissner effect, serves as one of the fundamental criterions to examine a candidate of superconductor \cite{Meissner_origin,RMP1999,RMP2018}. In order to justify a Meissner effect, one has to first observe a diamagnetic magnetization-temperature (M-T) curve with bifurcation between zero-field-cooling (ZFC) and field-cooling (FC) measurements, along with a superconducting hysteresis magnetization-magnetic field (M-H) loop below critical temperature ($T_{\rm c}$) with well-defined critical field ($H_{\rm c}$). The copper-substituted lead apatite (CSLA), also named as LK-99, has been claimed as a novel candidate for room-temperature superconductor \cite{Lee1,Lee2}, but a complete Meissner effect has not been reported up to date. Lee et al. reported a large diamagnetism, but it was stated to stem from Cu$_2$S as addressed by Habamahoro et al.\cite{2023habamahoro}. A more important hysteresis loop is still absent in the dc measurements\cite{Wang2023,Guo2023}  and has been merely observed in microwave circumstance \cite{2023strange}. There is no doubt the direct observation of dc hysteresis is essential, which turns out to be the main subject of this work.

To prevent ferromagnetism resulting from excessive copper doping, we devise and fabricate modified CSLA samples (Pb$_{9.1}$Cu$_{0.9}$(PO$_4$)$_6$S). The procedure involves a meticulous mixture of phosphate and lead sulfide in an aqueous solution through co-precipitation. Afterward, the mixture is heated under high pressure at 180~$^{\circ}$C for 24 hours, holding the solution with pH of 8. Following hydrothermal treatment, the samples are calcined under argon circumstance at 900~$^{\circ}$C for 8 hours. The temperature is then reduced to 500~$^{\circ}$C, and the calcination continues with a pure oxygen atmosphere for an additional 48 hours. Subsequently, the samples are allowed to cool to room temperature in presence of oxygen.

As reported previously \cite{2023strange}, by the state-of-the-art synthesis approach, the superconducting component in CSLA possesses pretty small scale, so the critical fields are as weak as several tens Oe. A strong paramagnetic signal may overwhelm possible low-field superconductivity, so the samples should be as pure as possible which may however greatly reduce the doping ratio of coppers and weaken the signals. More importantly, due to the robust memory effect of vortex glass phase, a sample that has been exposed to a strong magnetic field may also hold memory of the magnetization history. In this context, the measurement procedure of magnetic properties must be carefully designed and conducted.

We employed the MPMS-3 SQUID to conduct dc magnetization measurements on the samples. Manual positioning was utilized for measurement, with Fixed Center dc moment as data. To
illustrate the memory effect, a relaxed sample without initial magnetization was initiated to detect the ZFC curves at magnetic fields of 25~Oe and 200~Oe. Then we performed M-H curve measurements at temperatures of 300~K, 250~K, 200~K, and 100~K, respectively. After obtaining these curves, the sample underwent demagnetization to zero field. It was then cooled to 10~K and the ZFC-FC curve was again measured to determine the superconducting and glassy memory effect.

\begin{figure*}[t]
    \centering
    \includegraphics[width=0.7\linewidth]{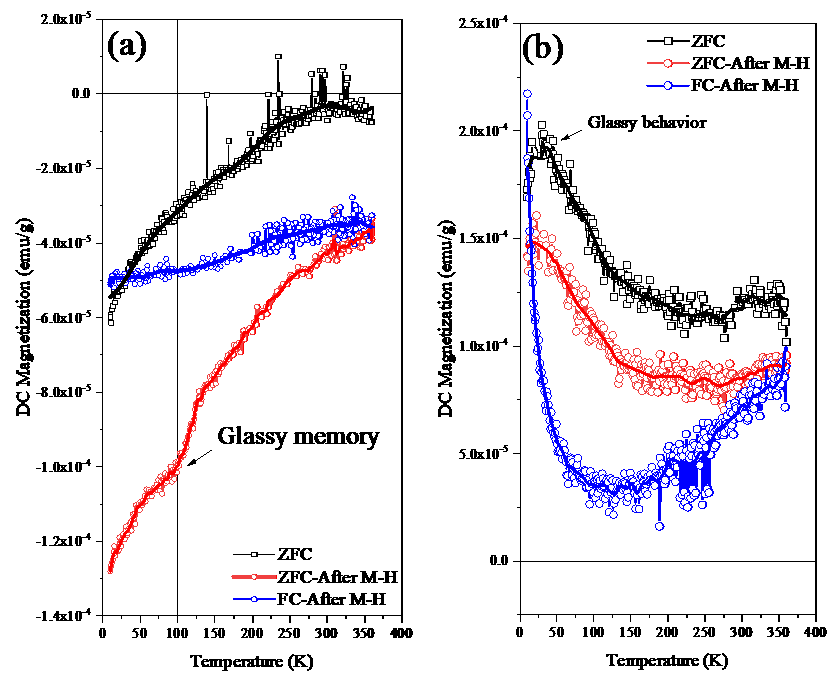}
    \caption{M-T curves with (a) 25~Oe and (b) 200~Oe. Black curves represent ZFC measurement before field sweep, while red and blue curves respectively refer to the ZFC and FC results after measuring M-H loops at 100K.}
    \label{fig1}
\end{figure*}

Fig.~1 shows the M-T curves before and after field sweep, which exhibits clear ZFC-FC bifurcation. All curves are diamagnetic under 25~Oe, while paramagnetism is present under 200~Oe, in agreement with the lower critical field $H_{\rm c1}$ being 30~Oe in low-field microwave absorption \cite{2023strange}. The ZFC curve after initial magnetization is lower than that before it, and there is an obvious kink at around 100~K, demonstrating the glassy memory effect as the field is finally swept at 100~K while cooling. There is also a turning point at around 250~K which can be recognized as the critical temperature $T_{\rm c}$. The glassy behavior is more observable under 200~Oe as the curves turn down below 50~K.

\begin{figure*}[t]
    \centering
    \includegraphics[width=0.7\linewidth]{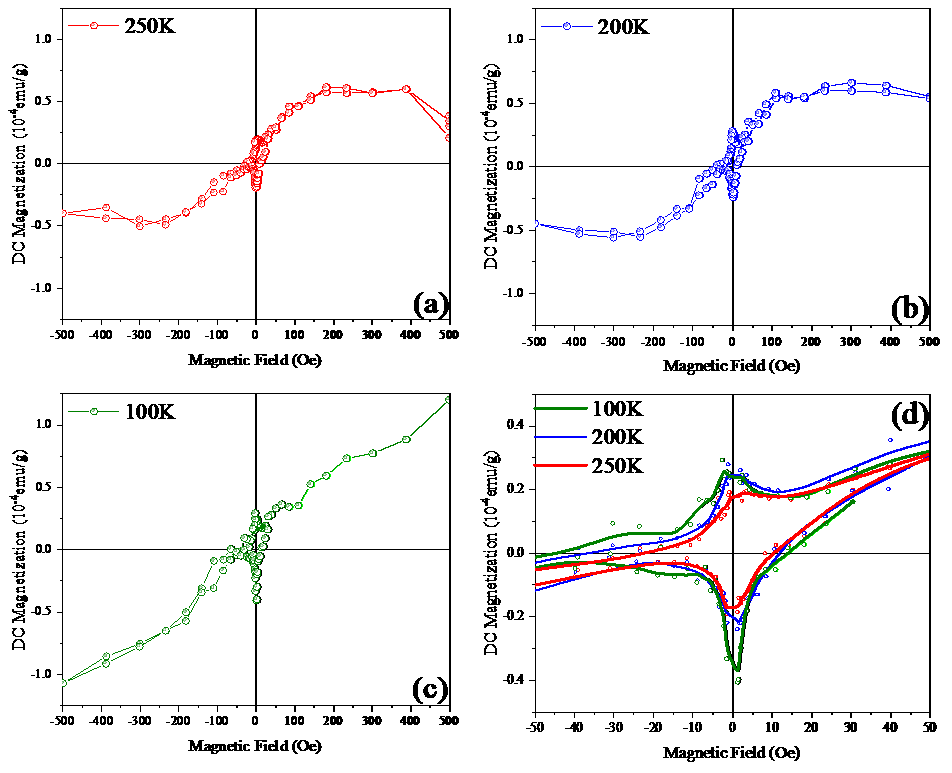}
    \caption{M-H hysteresis loops within $\pm500$~Oe at (a) 250K, (b) 200K, and (c) 100K, respectively. Detailed results within $\pm50$~Oe at different temperatures are shown in (d).}
    \label{fig1}
\end{figure*}

M-H curves at three temperatures are displayed in Fig.~2. Under strong magnetic field, the signals are basically paramagnetic. Below 10~Oe, typical superconducting hysteresis loops are clearly observed, although the signal noise ratios of the raw data are relatively small due to the tiny active portion of the sample. The hysteresis can not be recognized above 250~K. It is worth noting that, there is an asymmetry between forward and backward sweeps, that is, the negative peak under zero field is sharper than the positive one. This asymmetry has also been detected in the microwave absorption \cite{2023strange}. We guess this is because the first sweep of field is in the forward direction and the sample remembers it by generating relevant vortex currents, which has to be justified in the future studies.

\begin{figure*}[t]
    \centering
    \includegraphics[width=0.7\linewidth]{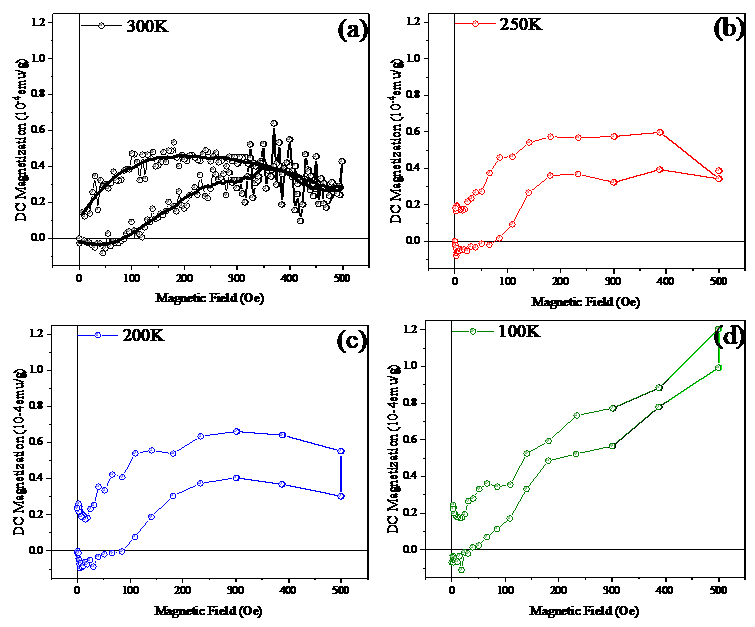}
    \caption{M-H curves of initial magnetization and first sweep backward at (a) 300K, (b) 250K (c) 200K and (d) 100K, respectively.}
    \label{fig1}
\end{figure*}

One may notice the magnetization at 25~Oe in the hysteresis loops are positive, different from that in the M-T curves. This suggests the initial magnetization curve is also important, so we separately display the initial magnetization curves and the first sweep-backward curves at different temperatures in Fig. 3. It is clear the magnetization is negative below 50~Oe in the initial curves which figures out a strange magnetization mechanism. There is also a hysteresis even at room temperature, and the bifurcation point is around 350~Oe. This hysteresis can also be seen in the microwave absorption, which in our opinion stems from the vortex glass phase \cite{RMP1994}. At lower temperatures, the bifurcation point increases and a peak appears at low field indicating the Meissner phase is possibly present.

\begin{figure*}[t]
    \centering
    \includegraphics[width=0.7\linewidth]{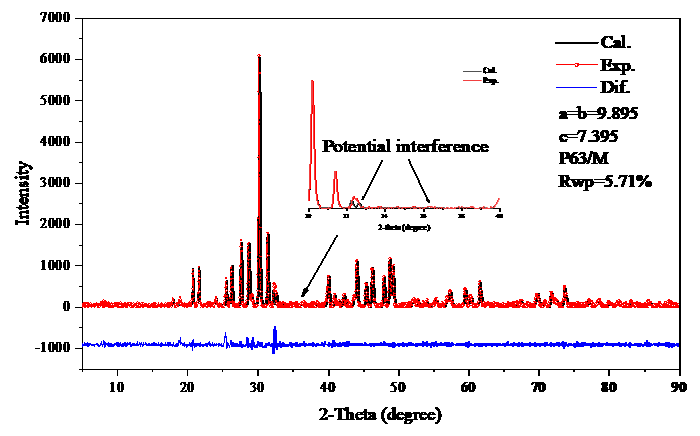}
    \caption{XRD spectrum of calculated, measured and difference between them. Inset shows the amplified range of 30--40$^{\circ}$.}
    \label{fig1}
\end{figure*}

The X-ray diffraction (XRD) data for the sample undergoes refinement using the Reflex module in Materials Studio, aligning with the P63/m structure characteristics of apatite, as illustrated in Fig.~4. However, a slight discrepancy is noted in the range of 25--27$^{\circ}$ and 30-40$^{\circ}$, potentially originated from a minor presence of residual oxide. Despite an extended period of roasting under a pure oxygen atmosphere, the interference from cuprous sulfide still persists, attributed to the intentional addition of sulfur elements during synthesis. Complete elimination of this interference remains challenging.

In summary, the diamagnetism in CSLA has been investigated via both M-T curves and hysteresis M-H loops, which can be observed up to 250~K. Given the ZFC-FC bifurcation at above 300~K, we think there is still great chance to observe room-temperature superconductivity. The signals in our sample are still extremely weak, so we have to devote efforts to further synthesizing scalable samples with more active components.

\section{Acknowledgments}

The authors gratefully acknowledge support from the National Natural Science Foundation of China (Grant Nos.~12374107, 11974118 and 52304430).

\section{Competing interests}

The authors declare no competing interests.

\bibliography{Meissner_v2.bbl}

\end{document}